\documentclass[10pt]{iopart}
\pdfoutput=1
\usepackage{graphicx}
\usepackage{amssymb}
\usepackage{xcolor}

\usepackage{iopams} 
\usepackage{listings}
\usepackage{float}

\eqnobysec
\newcommand{\beq}{\begin{equation}}
\newcommand{\beqa}{\begin{eqnarray}}
\newcommand{\eeq}{\end{equation}}
\newcommand{\eeqa}{\end{eqnarray}}

\renewcommand{\d}{{\rm d }}

\renewcommand{\max}{{\rm max}}
\newcommand{\prob}{\mathop{\rm Prob}\nolimits}

\renewcommand{\r}{{\rho}}

\newcommand{\stirl}[2]{\left(#1\atop#2\right)}

\newcommand{\fI}{f^{\rm I}}
\newcommand{\fII}{f^{\rm II}}
\newcommand{\fIII}{f^{\rm III}}
\newcommand{\LI}{\tau_{\max}^{\rm I}}
\newcommand{\LII}{\tau_{\max}^{\rm II}}
\newcommand{\LIII}{\tau_{\max}^{\rm III}}
\newcommand{\Q}{Q^{\star}}
\newcommand{\QI}{Q^{\rm I}}

\newcommand{\QIII}{Q^{\rm III}}

\renewcommand{\l}{\ell}

\newcommand{\lap}[1]{\mathrel{\mathop{\cal L}\limits_{#1}^{}}}

\newcommand{\zeq}[1]{\mathrel{\mathop{=}\limits_{#1}^{}}}
\newcommand{\zapprox}[1]{\mathrel{\mathop{\approx}\limits_{#1}^{}}}
\newcommand{\erf}{\mathop{\rm erf}}
\newcommand{\erfc}{\mathop{\rm erfc}}

\def\binom#1#2{{#1\choose #2}}
\newcommand{\yeqt}{y=t}
\newcommand{\deno}{U(t)}

\begin{document}

\title[Longest interval between zeros]{Longest interval between zeros of the tied-down random walk, the Brownian bridge and related renewal processes}
\date{\today}
\author{Claude Godr\`eche}
\address{
Institut de Physique Th\'eorique, Universit\'e Paris-Saclay, CEA and CNRS,
91191 Gif-sur-Yvette, France}\smallskip

\begin{abstract}
The probability distribution of the longest interval between two zeros of a simple random walk starting and ending at the origin, and of its continuum limit, the Brownian bridge, was analysed in the past by Ros\'en and Wendel, then extended by the latter to stable processes.
We recover and extend these results using simple concepts of renewal theory, which allows to revisit past or recent works of the physics literature.

\end{abstract}


\section{Introduction}

Problems that can be recast in the language of renewal processes appear recurrently in a number of studies of statistical physics without necessarily being recognized as such.
It is therefore useful to have access to this body of knowledge in simple terms.
This is one of the aims of the present study, where we revisit the question, investigated in the past by Wendel \cite{wendel}, of the longest interval between zeros of the Brownian bridge, seen as the continuum limit of the tied-down random walk (the simple random walk starting and ending at the origin), and of its generalization by a self-similar process with index $0<\theta<1$ (the Brownian bridge corresponding to $\theta=1/2$).
We recover his results using simple methods systematically developed in former studies of renewal processes \cite{gl2001,gms2015}.
We then extend this study in new directions.
Finally we use this knowledge to put several related works \cite{gms2015,fik,bar} in perspective.

The present study belongs to the more general area of extreme value statistics for intervals between events generated by a random process.
For zero crossings of the simple random walk or of Brownian motion, the intervals are the lengths of the excursions.
For points of a renewal process, the intervals are the ages of the components which are successively replaced.
For record times of a sequence of independent, identically distributed (iid) random variables, the intervals are the inter-record times or ages of the records.
For the record times of a random walk (the instants of time when the random walk reaches a maximum), the intervals are again the ages of the records.
While many studies have been devoted to the statistics of the longest interval for these examples \cite{lamperti,pitman,gourdon,csaki,gl2008,ziff,gms2009,gms2014,veto,review,gms2015}, much fewer have been devoted to this issue for the tied-down random walk, for the Brownian bridge, or for renewal processes constrained to have their last interval terminating at a given time \cite{wendel,pitman,fik,lindell}.
The purpose of the present work is to contribute to the investigation of these latter cases.

The detailed content of the paper is as follows.
In section \ref{sec:tiedRW} we start by recalling Ros\'en's results on the distribution of the longest interval between zeros of the tied-down random walk and its continuum limit, the Brownian bridge, as reported in \cite{wendel}.
We detail the derivation of this continuum limit and analyse the rescaled distributions of the longest interval and of its inverse, as well as their averages.
We then give the expression of the probability of a configuration of the tied-down walk in terms of the set of intervals between consecutive zeros and the number $M_N$ of these intervals, for a walk of $2N$ steps.
This in turn gives access to the marginal distribution of the number of intervals $M_N$ and provides an alternative derivation of the statistics of the longest interval.
It also yields the probability
that the last interval be the longest, or probability of record breaking.
In section \ref{sec:renewal} we give a reminder on renewal processes in continuous time, that is, with a continuous distribution of the lengths of the intervals between renewal events.
We then define the tied-down renewal process, which is a 
generalization of the renewal process in discrete time corresponding to the tied-down random walk (section \ref{sec:tied-down-renew}).
When the tail index of the distribution of intervals is less than 1, this process corresponds to the stable process considered in \cite{wendel}.
We analyse the rescaled distributions of the longest interval and of its inverse, as well as other characteristics of this process, such as the number of events, the statistics of a single interval and the probability of record breaking.
We finally consider the cases of a narrow distribution of intervals or a broad distribution with tail index 
$\theta>1$.
We close by revisiting past or recent relevant studies \cite{gms2015,fik,bar}.
Some definitions and derivations are relegated to the appendices.

\section{The tied-down random walk}
\label{sec:tiedRW}

\subsection{Ros\'en's results}
\label{sec:wendel}

\begin{figure}
\begin{center}
\includegraphics[angle=0,width=1.\linewidth]{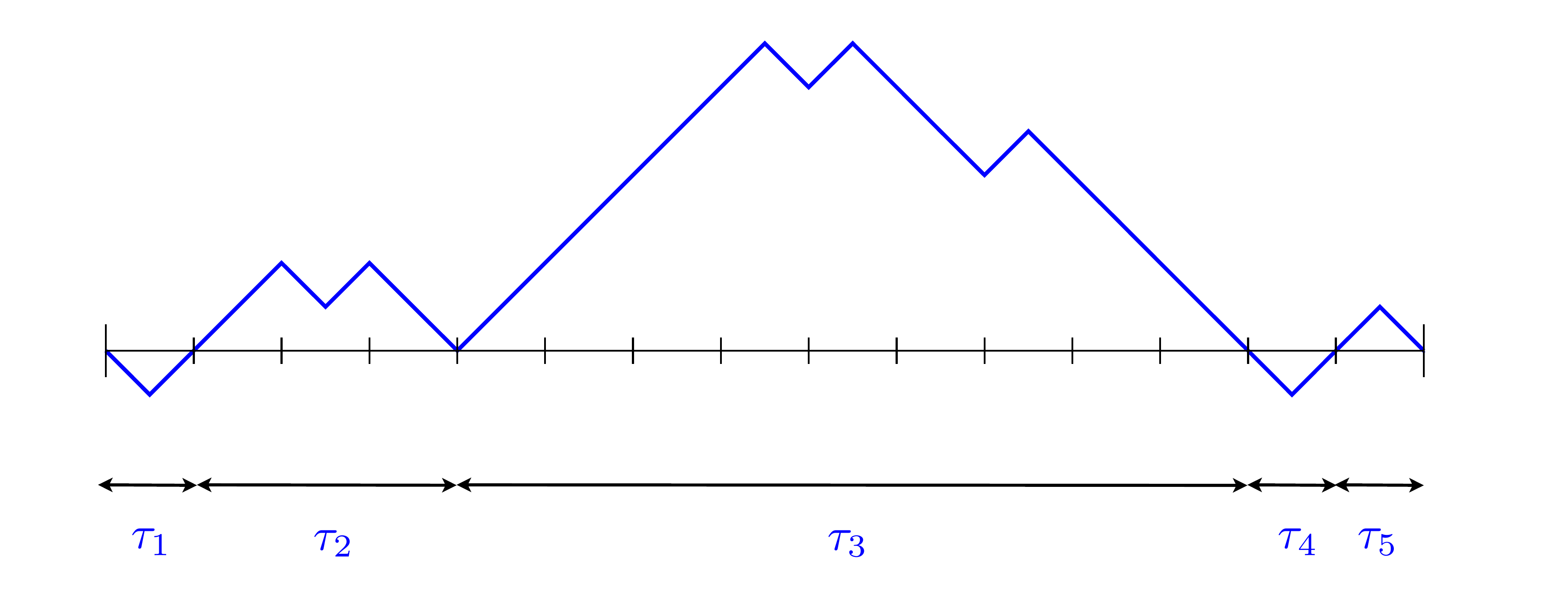}
\caption{A tied-down random walk, or Bernoulli bridge, is a simple random walk starting and ending at the origin.
In this example the walk is made of $2N=30$ steps, with $M_{15}=5$ intervals between zeros, $\tau_1,\dots,\tau_5$, taking the values $2,6,18,2,2$, respectively, and the longest interval $I_{15}\equiv\tau_3$.
The ticks on the $x-$axis correspond to two time-steps.
\label{fig:figure}}
\end{center}
\end{figure}

We first recall Ros\'en's results for the tied-down random walk as reported by Wendel in \cite{wendel}, keeping his notations.
Consider the sum $S_n$ of $n$ independent random variables taking the values $\pm1$, with probabilities $1/2$.
The random walk $S_n$ is conditioned to be `tied down' at time $2N$, i.e., to return to the origin at that time.
This walk is also named the Bernoulli bridge \cite{verwaat,labarbe}.
Let $I_N$ be the longest interval between consecutive zeros of $S_n$, during time $0\le n\le 2N$.
Define the joint probability
\beq
v_{N,k}=\prob(I_N\le 2k,S_{2N}=0),\qquad v_{0,k}=1.
\eeq
The quantity of interest is the conditional probability 
\beq\label{eq:coeur}
\prob(I_N\le 2k|S_{2N}=0)=\frac{v_{N,k}}{u_N},
\eeq
where $u_N$ is the probability of return of the walk at time $2N$ (see (\ref{eq:return}) in \ref{sec:return})
\beq\label{eq:uN}
u_N=\prob(S_{2N}=0)=v_{N,\infty}.
\eeq
The joint probability $v_{N,k}$ satisfies the renewal equation
\beq\label{eq:renewD}
v_{N,k}=\sum_{n=1}^{k}f_n\,v_{N-n,k},
\eeq
where $f_n$ is the probability of first return to zero at time $2n$ (see (\ref{eq:first}) in \ref{sec:return}).
From (\ref{eq:renewD}) we deduce that the generating functions 
\beq\label{eq:deffkz}
\tilde v_k(z)=\sum_{N\ge0}v_{N,k}z^N,\qquad \tilde f_k(z)=\sum_{n=1}^{k}f_n z^n
\eeq
are related by
\beq\label{eq:gf}
\tilde v_k(z)=\frac{1}{1-\tilde f_k(z)}.
\eeq

The result (\ref{eq:rosen1}) below, due to Ros\'en, as stated in \cite{wendel}, gives the continuum limit, at large times $2N$,
of the conditional probability (\ref{eq:coeur}), using (\ref{eq:gf}).
This conditional probability reads, using a star to indicate the tied-down condition,
\beq\label{eq:FR}
F_R^{\star}(r)=\lim_{N\to\infty}\prob(R_N\le r|S_{2N}=0)=\lim_{N\to\infty}\frac{v_{N,k=N r}}{u_N},
\eeq
also equal to
\beq\label{eq:FV}
\overline{F}_V^{\star}(v)=\lim_{N\to\infty}\prob(V_N>v|S_{2N}=0),
\eeq
the bar referring to the complementary distribution function,
with the following notations:
\beqa\label{eq:notations}
R_N=\frac{I_N}{2N},\qquad R^{\star}=\lim_{N\to\infty}R_N,\qquad r=\frac{k}{N}.
\nonumber\\
V_N=\frac{1}{R_N},\qquad V^{\star}=\frac{1}{R^{\star}},\qquad \hspace*{1.01cm} v=\frac{N}{k}=\frac{1}{r},
\eeqa
and where $r$ and $v$ are real variables, with $0<r<1$ and $v>1$.
According to \cite{wendel}, we have
\beqa\label{eq:rosen1}
F_R^{\star}(r)=
\overline{F}_V^{\star}(v)
&=&\pi\sqrt{v}\sum_{k=-\infty}^{\infty}(-2x_k)\e^{x_k(1+v)},
\\
&=&\pi \sqrt{v}\,\fII_V(v),
\label{eq:rosen2}
\eeqa
denoting by $\fII_V(v)$ the sum on the right side of (\ref{eq:rosen1}), 
and where the $x_k$ are the zeros of the function
\beq
D(x)=1+\sqrt{\pi x}\,\e^{x}\erf \sqrt{x},\qquad \erf x=\frac{2}{\sqrt{\pi}}\int_0^x {\rm d}u\,\e^{-u^2}.
\eeq
The error function $\erf x$ being odd in $x$, the function $\sqrt{x}\erf \sqrt{x}$ is entire and has only zeros in the complex $x$ plane.
These zeros have all negative real parts, for instance
\beq
\fl x_0=-0.854\dots,x_{\pm1}=-4.248\dots\pm {\rm i\, } 6.383\dots,
x_{\pm2}=-5.184\dots\pm {\rm i\, } 12.885\dots
\eeq
and so on.
Let us note that 
the Laplace transform of $\fII_V(v)$ with respect to $v$ is given by
\beq\label{eq:fhatII}
\widehat{\fII_V}(x)
=\frac{\e^{x}}{1+\sqrt{\pi x}\,\e^{x}\erf\sqrt{x}},
\eeq
where the variable $x$ is conjugate to $v$.
This can be seen by taking the inverse Laplace transform of $\widehat{\fII_V}(x)$ and noting that
the residues of this function at the poles $x_k$ are equal to $-2 x_k\,\e^{x_k}$ (see (\ref{eq:eqdiff}) with $\theta=1/2$ in section \ref{sec:charact}), which yields $\fII_V(v)$ back.
We can now interpret $\fII_V(v)$ as the density of a random variable, because it is a positive function, normalised to unity since $\widehat{\fII_V}(0)=1$.
The precise meaning of this density is given in section \ref{sec:longest}.
Let us note that the function $D(x)$ is equal to the confluent hypergeometric function $_1 F_1(1,1/2,x)$ (see (\ref{eq:fIx})).

\subsection{Proof of (\ref{eq:rosen1})}

Let us now detail how to derive the continuum scaling limit (\ref{eq:rosen1}) from (\ref{eq:gf}),
by an asymptotic analysis of the latter at large times.
Since $f_n=u_{n-1}-u_n$ (see \ref{sec:return}),
we have
\beq\label{eq:parties}
1-\tilde f_k(z)=u_kz^k+(1-z)\sum_{n=0}^{k-1}u_n z^n,
\eeq
then setting $z=\e^{-s}$ and using (\ref{eq:asymu}), we obtain, when $k\to\infty$, $s\to0$, with $x=ks$ fixed, 
\beqa
\tilde v_k(z)=\frac{1}{1-\tilde f_k(z)}
&\approx&\frac{1}{\sqrt{s}}\frac{1}{(\pi ks)^{-1/2}\e^{-ks}+\erf \sqrt{ks}}
\nonumber\\
&\approx&\sqrt{\pi k}\,\frac{\e^{x}}{1+\sqrt{\pi x}\,\e^{x}\erf \sqrt{x}}.
\label{eq:scale}
\eeqa
It is now simple to infer (\ref{eq:rosen1}) from (\ref{eq:scale}).
In the continuum scaling limit (\ref{eq:notations}),
informally denoting the generating function $\tilde v_k(z)$ as a Laplace transform with respect to $N$ (considered now as a real variable), yields
\beqa
\fl\tilde v_k(z)
=\lap{N}\prob\left(R_N\le k/N,S_{2N}=0\right)
=\lap{N}\prob\left(V_N> N/k,S_{2N}=0\right).
\eeqa
Rescaling $N$ by $k$, the left side becomes the Laplace transform with respect to $v$, with $x$ conjugate to $v$, 
\beq
 \lap{v}\prob(V_N> v=N/k,S_{2N}=0)
\approx\sqrt{\frac{\pi}{k}}\widehat{\fII_V}(x),
\eeq
hence
\beq
\prob(V_N> v,S_{2N}=0)
\approx\sqrt{\frac{\pi}{k}}\fII_V(v).
\eeq
Dividing both sides by $u_N\approx1/\sqrt{\pi N}$ leads to (\ref{eq:rosen1}).

\subsection{Characterization of the density}
\label{sec:wendelfstar}

\begin{figure}
\begin{center}
\includegraphics[angle=0,width=1\linewidth]{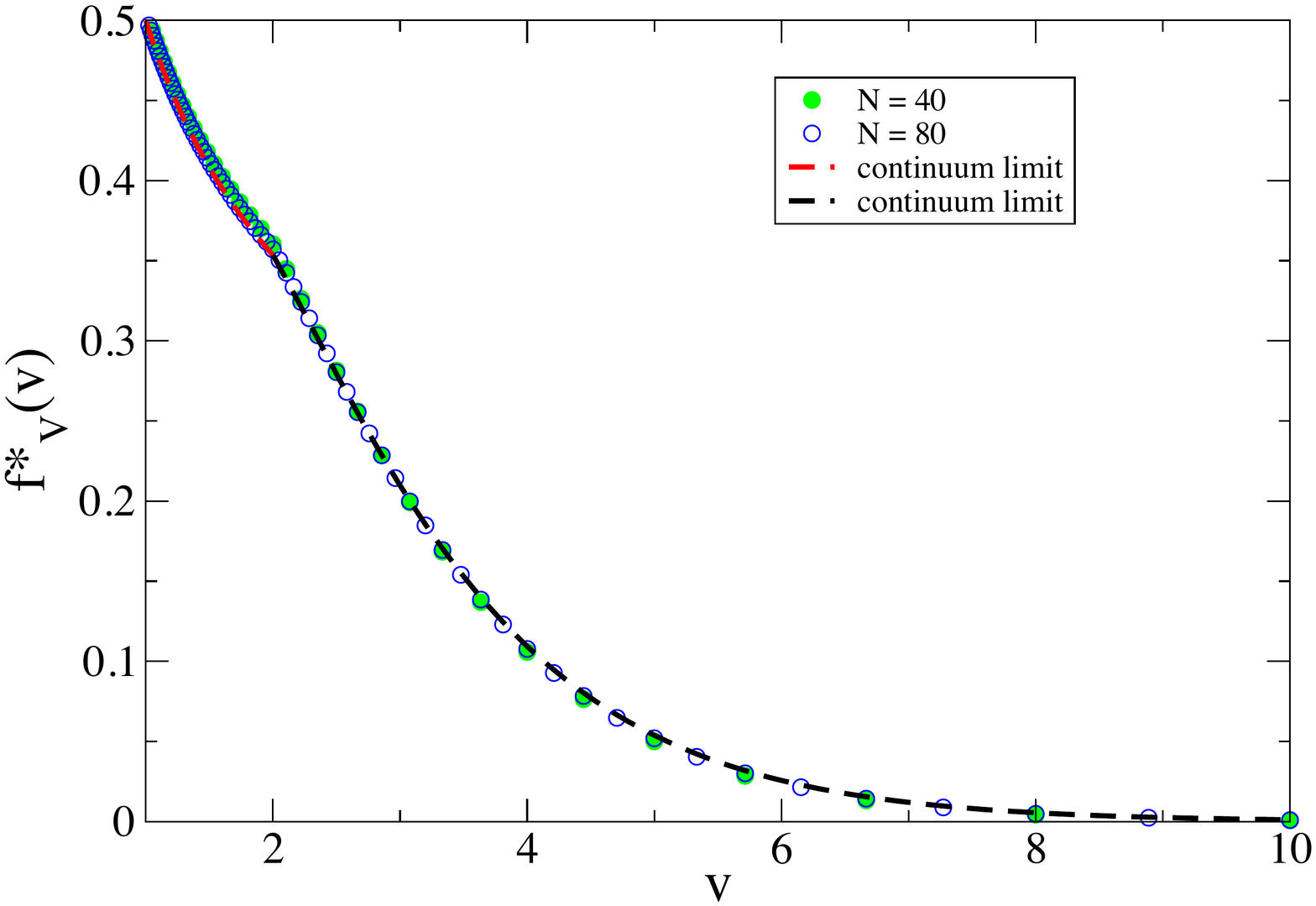}
\caption{Probability density $f_V^{\star}(v)$ for the tied-down random walk obtained from (\ref{eq:gf}) for $N=40$ and $N=80$.
The black dashed curve ($v>2$) corresponds to (\ref{eq:fVas}),
the red dashed curve ($1<v<2$) to (\ref{eq:fV12}).
\label{fig:fVstar}}
\end{center}
\end{figure}
\begin{figure}
\begin{center}
\includegraphics[angle=0,width=1\linewidth]{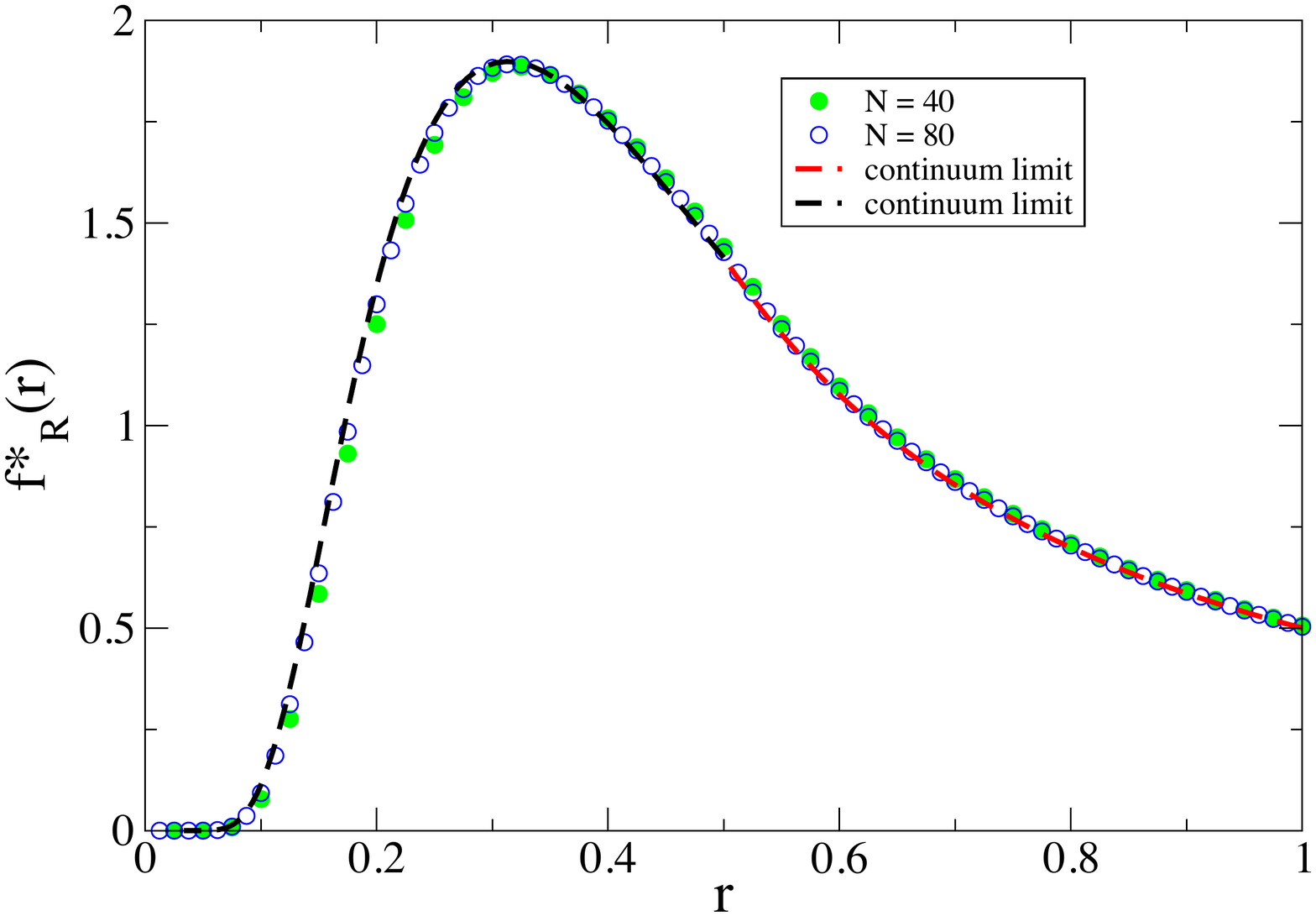}
\caption{Probability density $f_R^{\star}(r)$ obtained from the data of figure \ref{fig:fVstar}.
The black dashed curve ($r<1/2$) corresponds to (\ref{eq:fRas}),
the red dashed curve ($1/2<r<1$) to (\ref{eq:fR12}).
}
\label{fig:fRstar}
\end{center}
\end{figure}

The large $v$ behaviour of the density $f^{\star}_V(v)$ can be read off from (\ref{eq:rosen1}).
At leading order 
\beq\label{eq:FVas}
\overline{F}^{\star}_V(v)\approx 2|x_0|\pi\sqrt{v}\,\e^{-|x_0|(1+v)},
\eeq
from which $f^{\star}_V(v)$ ensues by derivation,
\beq\label{eq:fVas}
f^{\star}_V(v)\approx 2\pi\, x_0^2\sqrt{v}\,\e^{-|x_0|(1+v)}\left(1-\frac{1}{2|x_0|v}\right).
\eeq
As a consequence, the density $f^{\star}_R(r)=\d F^{\star}_R(r)/\d r $ has an essential singularity at the origin,
\beq\label{eq:fRas}
f^{\star}_R(r)\approx 2\pi\, x_0^2\frac{\e^{-|x_0|(1+1/r)}}{r^{5/2}}\left(1-\frac{r}{2|x_0|}\right).
\eeq
The density $f^{\star}_V(v)$ is a piece-wise continuous function \cite{wendel}, as is also the case of the densities $\fI_V(v)$, $\fII_V(v)$ and $\fIII_V(v)$ \cite{gms2015,lamperti} ($\fI_V$ and $\fIII_V$ are defined later).
The behaviour for $1< v< 2$ is given in \cite{wendel}:
\beq\label{eq:fV12}
f^{\star}_V(v)=\frac{1}{2\sqrt{v}},
\eeq
hence, for $1/2< r< 1$,
\beq\label{eq:fR12}
f^{\star}_R(r)=r^{-3/2}/2.
\eeq
The reasoning is again due to Ros\'en.
Consider
\beq
w_{N,k}=\prob(I_N=2k,S_{2N}=0).
\eeq
When $I_N=2k>N$, then the longest interval is unique.
This is the case in figure \ref{fig:figure}, where $I_{15}=18$.
Decomposing a path into three contributions, we obtain
\beq\label{eq:rosenbis}
w_{N,k}=\sum_{n=0}^{N-k}u_n\,f_k\,u_{N-n-k}=f_k, \quad (2k>N),
\eeq
the last equality resulting from the identity $\sum_{j=0}^{m}u_ju_{m-j}=1$.
In figure \ref{fig:figure} these three contributions correspond to $\tau_1+\tau_2$, $\tau_3$, and $\tau_4+\tau_5$.
So, using (\ref{eq:asymu}), we have
\beqa
F^{\star}_R(r)
&=&1-\lim_{N\to\infty}\sum_{k=Nr}^N
\frac{w_{N,k}}{u_N}
\\
&=&1-\int_{r}^1 {\rm d}y \frac{1}{2y^{3/2}}=2-\frac{1}{r^{1/2}},\quad (1/2<r<1).
\eeqa
This result will be generalized in section \ref{sec:charact}.
A method for the determination of $f^{\star}_V(v)$ in the successive intervals $(i,i+1)$ is given in \cite{wendel}.
Complementary information on this issue can be found in \cite{fik,pitman,gourdon,lindell}.

Figure \ref{fig:fVstar} depicts the density $f_V^{\star}(v)$ obtained by extracting $v_{N,k}$ from (\ref{eq:gf}) by formal computation, with $0\le k\le N$, for $N=40$ and $N=80$, then rescaling appropriately this sequence.
The discontinuity of the derivative at $v=2$ is clearly visible.
Figure \ref{fig:fRstar} depicts the density $f_R^{\star}(r)$ obtained from the same data.
For both curves the scaling is already good for these rather small values of $N$.
The dashed curves in figure \ref{fig:fVstar} correspond to the predictions (\ref{eq:fVas}) and (\ref{eq:fV12}), and to 
the predictions (\ref{eq:fRas}) and (\ref{eq:fR12}) in figure \ref{fig:fRstar}.
The good adequation of the asymptotic predictions (\ref{eq:fVas}) and (\ref{eq:fRas}) with the scaling curves in all the domains $v>2$ and $r<1/2$ is due to the large gap between the values of $x_0$ and of the real part of $x_1$.

\subsection{Average longest interval}
\label{sec:averageRW}

Since
\beq
\prob(I_N\le 2k,S_{2N}=0)+\prob(I_N> 2k,S_{2N}=0)=u_N,
\eeq
the generating function of the sum adds to $1/\sqrt{1-z}$.
In the continuum limit we thus find, using (\ref{eq:scale}), that
\beq
\lap{N}\prob(I_N> 2k,S_{2N}=0)\approx\frac{1}{\sqrt{s}}
\frac{1-\sqrt{\pi x}\e^x \erfc \sqrt{x}}{1+\sqrt{\pi x}\e^x \erf \sqrt{x}},
\eeq
where $\erfc x=1-\erf x$.
Defining
\beq\label{eq:deffVIII}
\widehat{\fIII_V}(x)=\frac{1-\sqrt{\pi x}\,\e^x \erfc \sqrt{x}}{1+\sqrt{\pi x}\,\e^x \erf \sqrt{x}}
=1-\sqrt{\pi x}\,\widehat{\fII_V}(x),
\eeq
whose meaning is given in section \ref{sec:longest}, we have
\beqa
\lap{N}\langle I_N,S_{2N}=0\rangle
&=&\lap{N}\int_0^\infty{\rm d}(2k)\,
\prob(I_N> 2k,S_{2N}=0)
\nonumber\\
&=&\frac{2}{s^{3/2}}\int_0^\infty{\rm d}x\,\widehat{\fIII_V}(x)= \frac{2}{s^{3/2}}0.2417\ldots.
\eeqa
By inversion of the Laplace transform and division by $u_N$, we obtain
\beq
\langle I_N|S_{2N}=0\rangle\approx 4N\times 0.2417\ldots.
\eeq
and finally
\beq\label{eq:averageRW}
\langle R^{\star}\rangle=\lim_{N\to\infty}\langle R_N|S_{2N}=0\rangle= 0.4834\ldots.
\eeq
This constant is also mentioned for related problems in \cite{gourdon}.
We shall comment and generalize this result later (see (\ref{eq:Rmoyen})).

\subsection{Probability of a configuration for the tied-down random walk}
\label{sec:renew-discret}

An alternate method to recover the results above consists in considering the probability of a configuration of the walk, in terms of the successive intervals $\tau_1,\tau_2,\dots$ between zeros.
Moreover this allows to investigate new quantities, such as the number of intervals up to time $2N$, or the probability of record breaking, i.e., the probability that the last interval be the longest.
This formalism will also serve as a preparation for the sequel, where we consider the continuum renewal process generalizing the case of the random walk.

Let $M_N$ be the number of intervals up to time $2N$.
This random variable takes the values $m=0,1,2,\dots$.
(For the tied-down walk, $m=0$ necessarily implies $2N=0$.)
A configuration of the tied-down walk is specified by $\{\tau_1, \dots,\tau_{M_N},M_{N}\}$, whose realization is denoted by
$\{2\l_1,\dots,2\l_{m},m\}$, with $\sum_i\l_i=N$.
These definitions are illustrated in figure \ref{fig:figure}.
The probability of this configuration is 
\beq\label{eq:Pstar}
\fl\prob(\tau_1=2\l_1,\dots,\tau_{M_N}=2\l_{m},M_N=m|S_{2N}=0)=\frac{f_{\l_1}\dots f_{\l_m}\delta(\sum_{i=1}^m\l_i,N)}{u_N}.
\eeq
The numerator in the right side of the equation expresses the fact that the walk starts afresh each time it crosses the origin and therefore that the time intervals $\tau_i$ have the common distribution $\prob(\tau=2\l)\equiv f_\l$, with the constraint that they sum to the total time $2N$, which is ensured by the presence of the Kronecker delta, $\delta(i,j)=1$ if $i=j$ and $0$ otherwise.
The time intervals $\tau_i$ thus form a discrete renewal process (see section \ref{sec:renewal} for the general definition of a renewal process).
The denominator, obtained from the numerator by summing on the $\l_i\ge1$ and $m$, is precisely the probability $u_N$ of return of the walk at time $2N$, defined in (\ref{eq:return}),
\beq\label{eq:uN-renew}
u_N=\sum_{m\ge0}\sum_{\l_1\dots\l_m}f_{\l_1}\dots f_{\l_m}\delta\Big(\sum_{i=1}^m\l_i,N\Big).
\eeq
This can be checked 
by taking the generating function of the right side of (\ref{eq:uN-renew})
\beqa
\sum_{N\ge0}z^N\,\sum_{m\ge0}\sum_{\l_1\dots\l_m}f_{\l_1}\dots f_{\l_m}\delta\Big(\sum_{i=1}^m\l_i,N\Big)
=\sum_{m\ge0}\tilde f(z)^{m}
\\\label{eq:uzfz}
=\frac{1}{1-\tilde f(z)}=\tilde u(z),
\eeqa
which is indeed the generating function of the $u_N$.
This can be alternatively be checked as follows.
Let us denote the sum of the random number $M_N$ of intervals by
\beq
t_{M_N}=\tau_1+\cdots+\tau_{M_N},
\eeq
and the corresponding sum when $M_N$ is fixed equal to $m$ by
\beq
t_m=\tau_1+\cdots+\tau_{m}.
\eeq
Then, clearly,
\beqa\label{eq:probtm}
\prob(t_{m}=2N)&=&\prob(M_N=m,S_{2N}=0)
\nonumber\\
&=&\sum_{\l_1\dots\l_m}f_{\l_1}\dots f_{\l_m}\delta\Big(\sum_{i=1}^m\l_i,N\Big).
\eeqa
Hence, summing on $m$,
\beq\label{eq:uNbis}
\sum_{m\ge0}\prob(t_{m}=2N)=
\prob(t_{M_N}=2N)\equiv u_N.
\eeq

We can now easily express the probability distribution of the number of intervals (in the tied-down case) as
\beqa\label{eq:pstarm}
p^{\star}_m(N)=\prob(M_N=m|S_{2N}=0)=\frac{\prob(t_{m}=2N)}{u_N}
\\
=\frac{\left[\tilde f(z)^m\right]_{N}}{u_N},
\eeqa
where the notation $[\cdot]_N$ means the $N-$th coefficient of the series inside the brackets.
In particular, the mean number of intervals is
\beq
\langle M_N|S_{2N}=0\rangle=\sum_{m\ge0}m\, p^{\star}_m(N).
\eeq
The generating function of the numerator of this expression reads
\beq
\sum_{m\ge0}m\tilde f(z)^m=\frac{\tilde f(z)}{\big(1-\tilde f(z)\big)^2}.
\eeq
This yields, using (\ref{app:ftilde}),
\beq\label{eq:MNdiscret}
\langle M_N|S_{2N}=0\rangle=\frac{1}{u_N}-1=\frac{2^{2N}}{\binom{2N}{N}}-1.
\eeq
Hence, at long times $2N$, 
\beq\label{eq:meanMN}
\langle M_N|S_{2N}=0\rangle\approx\sqrt{\pi N}.
\eeq
More generally, in the continuum limit where the series in the expressions above are dominated by $z$ close to 1, the distribution $p^{\star}_m(N)$ has the scaling form
\beq\label{eq:pmN}
p^{\star}_m(N)\approx \frac{1}{2\sqrt{N}}y\,\e^{-y^2/4},\qquad y=\frac{m}{\sqrt{N}},
\eeq
from which the asymptotic expressions of the higher moments ensue,
\beq
\langle (M_N)^p|S_{2N}=0\rangle=2^{p-1}N^{p/2}p\,\Gamma\left(\frac{p}{2}\right).
\eeq
The meaning of (\ref{eq:meanMN}) is that the typical interval $\tau_i$ has length of order $\sqrt{N}$, to be compared to the length of order $N$ of the longest interval.
This will be generalized below (see (\ref{eq:single})). 

\subsubsection*{The longest interval.}

Knowing (\ref{eq:Pstar}) yields immediately
the conditional probability (\ref{eq:coeur})
\beqa\label{eq:problongest}
\prob(I_N\le 2k|S_{2N}=0)=\frac{v_{N,k}}{u_N}
\nonumber\\
=\frac{1}{u_N}
\sum_{m\ge0}\sum_{\l_1=1}^{k}\dots\sum_{\l_{m}=1}^{k}f_{\l_1}\dots f_{\l_m}\delta\Big(\sum_{i=1}^m\l_i,N\Big).
\eeqa
Taking the generating function of the numerator leads again to (\ref{eq:gf}).

\subsection{Probability that the last interval is the longest }
\label{sec:breaking}
When dealing with the successive extremes, or records, taken by a series of random variables, an important indicator of the statistics of records is the probability that the last random variable is the largest, named the probability of record breaking.
For independent, identically distributed (iid) random variables this probability is equal to the inverse of the number of random variables in the series \cite{renyi,arnold}.
For correlated random variables this quantity scales differently.
For instance, when considering the intervals between the record times of iid random variables, this probability converges asymptotically to a constant \cite{gl2008}.
The same holds for the intervals between renewal points when the distribution of intervals has a tail exponent $\theta$ (as defined in section \ref{sec:renewal}) less than one \cite{gms2009,gms2015,gms2014} (this corresponds to cases I and II in section \ref{sec:jointf}).
However if, in the sequence of intervals, one discards the last unfinished one
(this corresponds to case III in section \ref{sec:jointf}),
then the probability that the last interval is the longest one scales as $t^{-\theta}$ up to logarithmic corrections \cite{gms2015,gms2014}, i.e., as $1/\sqrt{N}$ up to logarithmic corrections for the excursions of a random walk.
We are thus led to investigate this quantity for the tied-down random walk and later for the tied-down renewal process in order to compare with the known unconstrained cases.

The determination of the probability of record breaking for the tied-down random walk, i.e.,
\beq\label{eq:QN}
Q^{\star}_N=\prob(I_N=\tau_{M_N}|S_{2N}=0),
\eeq
proceeds as follows (see \cite{gms2014} for similar reasonings).
We have
\beq
 Q^{\star}_N=\sum_{m\ge1}\prob(I_N=\tau_{M_N},M_N=m|S_{2N}=0)=\frac{Q_N}{u_N},
\eeq
where $Q_N=\sum_{m\ge1}Q_{N,m}$, and
\beq
Q_{N,m}=\sum_{\l_m\ge1}\sum_{\l_1=1}^{\l_m}\dots\sum_{\l_{m-1}=1}^{\l_m}f_{\l_1}\dots f_{\l_m}\delta\Big(\sum_{i=1}^m\l_i,N\Big).
\eeq
Taking the generating function of this quantity and summing on $m$ we obtain
\beq
\tilde Q(z)=\sum_{N\ge0} Q_N z^N=\sum_{\l\ge1}\frac{f_\l z^\l}{1-\tilde f_\l(z)},
\eeq
where $\tilde f_\l(z)$ is defined in (\ref{eq:deffkz}).
This quantity appears in the appendix of \cite{gms2014},
where it is found that, for $z\to1$, 
\beq
\tilde Q(z)\approx -\ln \sqrt{1-z}+c, \qquad c=\frac{1}{2}\Big(\gamma+\ln\frac{4}{\pi}\Big)\approx 0.409,
\eeq
($\gamma$ is the Euler constant).
By inversion and division by $u_N$, we thus finally obtain
\beq\label{eq:QN+}
Q^{\star}_N\approx \frac{\sqrt{\pi}}{2\sqrt{N}}.
\eeq
This result can be interpreted as follows.
As recalled above, for iid random variables the probability of record breaking is equal to the inverse of the number of variables.
The same holds if the random variables are exchangeable.
In the present case, our intuition is that the intervals $\tau_1,\dots,\tau_{M_N}$ all play the same role.
So we are led to compute the average of the inverse of the number of intervals $M_N$.
Using (\ref{eq:pstarm}), we have
\beq
\left\langle\frac{1}{M_N}\right\rangle=\sum_{m\ge1} \frac{p^{\star}_m(N)}{m}
=\frac{\left[-\ln\sqrt{1-z}\right]_{N}}{u_N}
=\frac{1}{2Nu_N}.
\eeq
This expression is indeed asymptotically equal to (\ref{eq:QN+}).
As we shall see in section \ref{sec:Qt}, the same holds, not only asymptotically, but also at any time for the corresponding continuum renewal process (see (\ref{eq:Qstar})).

\textcolor{black}{All the formalism seen in this section can be generalized to the case of an arbitrary distribution of intervals $f_{\l}$.
The process thus defined is just the counterpart in discrete time of the tied-down renewal process of section \ref{sec:tied-down-renew}.
}
\section{A reminder on renewal processes}
\label{sec:renewal}

As mentioned in the previous section, the walk starts afresh each time it crosses the origin.
This allows to describe the statistical properties of the intervals between zeros of the tied-down walk in terms of a renewal process in discrete time
yielding the expression of the conditional probability of a configuration (\ref{eq:Pstar}).
The present section and the next one generalize these properties to a renewal process in continuous time, with an arbitrary distribution of intervals \cite{feller,doob,cox}.
In particular, the generalization of (\ref{eq:Pstar}) is given by (\ref{eq:fstar+})
and the results of section \ref{sec:tiedRW} corresponding to Brownian motion are recovered by taking $\theta=1/2$ in what follows.
We refer to table \ref{tab:table} for the correspondences between the notations for the discrete random walk and the continuum renewal process presented below.
\begin{table}[h]
\caption{Correspondences between the notations for the discrete random walk of section \ref{sec:tiedRW} and the continuum renewal process of sections \ref{sec:renewal} and \ref{sec:tied-down-renew}.}
\label{tab:table}
\begin{center}
\begin{tabular}{|c|c|}
\hline
 Random walk & Renewal proc. \\
\hline
$2N$&$t$\\
$2k$&$\l$\\
$I_N$&$\tau_{\max}(t)$\\
$u_N$&$\deno$\\
$v_{N,k}$&$F(t;\l)$\\
$f_n$&$\rho(\tau)$\\
$M_N$&$N_t$\\
$\prob(t_m=2N)$&$f_{t_n}(t)$\\
$w_{N,k}$&$w(t;\l)$\\
\hline
\end{tabular}
\end{center}
\end{table}

\subsection{Definitions and observables}
We start by reminding the definitions and notations used for renewal processes, following \cite{gl2001}.
Events occur at the random epochs of
time $t_{1},t_{2},\ldots$, from some time origin $t=0$. 
These events are for instance the zero crossings of some stochastic process, (or zeros as for the simple random walk of section \ref{sec:wendel}).
We take the origin of time on a zero crossing. 
When the intervals of time between events, $\tau
_{1}=t_{1},\tau_{2}=t_{2}-t_{1},\ldots $, are independent and identically
distributed random variables with common density $\rho (\tau)$, the process thus
formed is a \textit{renewal process} \cite{feller,doob,cox,lamperti}. 
A renewal process is thus a simple generalisation of the Poisson process.

The probability $p_0(t)$ that no event occurred up to time $t$
is simply given by the
tail probability: 
\beq
p_{0}(t)=\prob(\tau_1 >t)=\int_{t}^{\infty }{\rm d}\tau\,\rho (\tau). 
\label{tail}
\eeq
The density $\rho(\tau)$ can be either a narrow distribution with all
moments finite, in which case the decay of $p_{0}(t)$, as $t\rightarrow \infty $,
is faster than any power law, or a distribution characterized by a
power-law fall-off with index $\theta>0 $ 
\beq\label{eq:p0}
p_0(t)=\int_{t}^{\infty }{\rm d}\tau\,\rho (\tau)
\approx \left( \frac{\tau_{0}}{t}\right) ^{\theta }, 
\eeq
where $\tau_{0}$ is a microscopic time scale. 
If $\theta <1$ all moments of 
$\rho(\tau) $ are divergent, if $1<\theta <2$, the first moment 
$\left\langle \tau\right\rangle $ is finite but higher moments are
divergent, and so on. 
In Laplace space, where $s$ is conjugate to $\tau$, for a narrow
distribution we have 
\begin{equation}
\fl \lap{\tau}\rho (\tau)=\hat{\rho}(s)=\int_0^\infty{\rm d}\tau\, \e^{-s \tau}\rho(\tau)
\zeq{ s\rightarrow 0}
1-\left\langle 
\tau\right\rangle s+\frac{1}{2}
\left\langle \tau^{2}\right\rangle s^{2}+\cdots \label{ro_narrow}
\end{equation}
For a broad distribution, (\ref{eq:p0}) yields 
\begin{equation}
\hat{\rho}(s)\zapprox{s\rightarrow 0}\left\{ 
\begin{array}{ll}
1-a\,s^{\theta }, & (\theta <1) \\ 
1-\left\langle \tau\right\rangle s+a\,s^{\theta }, & (1<\theta <2),
\end{array}
\right. \qquad \label{ro_broad}
\end{equation}
and so on, where 
\beq\label{eq:a}
a=|\Gamma (1-\theta )|\tau_{0}^{\theta }.
\eeq
From now on, unless otherwise stated, we shall only consider the case $0<\theta<1$.

The quantities naturally associated to a renewal process \cite{feller,cox,gl2001} are the following.
The number of events which occurred between $0$ and $t$, i.e., the largest $n$ such that $t_n\le t$, is a random variable denoted by $N_t$.
The time of occurrence of the last event before $t$, that is of the 
$N_t-$th event, is therefore the sum of a random number of random variables denoted by
\beq\label{eq:tN}
t_{N}=\tau_{1}+\cdots +\tau_{N}. 
\eeq
The backward recurrence time $A_{t}$ 
is defined as the length of time
measured backwards from $t$ to the last event before $t$, i.e., 
\beq
A_{t}=t-t_{N}.
\eeq
It is therefore the age of the current, unfinished, interval at time $t$.
Finally the forward recurrence time (or excess time) $E_{t}$ is the time
interval between $t$ and the next event 
\beq
E_{t}=t_{N+1}-t. 
\eeq
We have the simple relation $A_t+E_t=t_{N+1}-t_{N}=\tau_{N+1}$.

\subsection{Three possible definitions for the last interval}
\label{sec:jointf}

Consider the following sequences of intervals \cite{gms2015}
\beqa\label{eq:sequences}
({\rm I}):\ &&\{\tau_1, \tau_2, \ldots, \tau_{N}, A_t \} ,
\nonumber\\
({\rm II}):\ &&\{\tau_1, \tau_2, \ldots, \tau_{N}, \tau_{N+1} \}, 
\nonumber\\
({\rm III}):\ &&\{\tau_1, \tau_2, \ldots, \tau_{N}\}.
\eeqa
To each of these sequences, supplemented by $N_t$, is associated a joint probability density \cite{gl2001,gms2015}.
For the first sequence, this joint probability density is, with the notations of \ref{sec:notations},
\beq\label{eq:fIjoint}
\fl f_{\vec{\tau},A_t,N_t}(t;\l_1,\ldots,\l_n,a,n)
=\r(\l_1)\ldots\r(\l_n)\,p_0(a)\,\delta\Big(\sum_{i=1}^n\l_i+a-t\Big).
\label{eq:fI}
\eeq
Likewise, the joint probability density of $\tau_1,\ldots,\tau_{N+1},N_t$ is
\beqa
\fl f_{\vec{\tau},\tau_{N+1},N_t}(t;\l_1,\ldots,\l_{n+1},n)
=\r(\l_1)\ldots\r(\l_{n+1})\,I(t_n<t<t_{n}+\l_{n+1}),
\label{eq:fII}
\eeqa
where $I(\cdot)=1$ or $0$ if the condition inside the parentheses is satisfied or not.
Finally, for the third sequence, the joint probability density of $\tau_1,\ldots,\tau_{N},N_t$ is
\beqa\label{eq:fIII}
\fl f_{\vec{\tau},N_t}(t;\l_1,\ldots,\l_n,n)
=\r(\l_1)\ldots\r(\l_n)\,\int_0^\infty{\rm d}a\,p_0(a)\,\delta\left(\sum_{i=1}^n\l_i+a-t\right),
\eeqa
which can alternatively be obtained from (\ref{eq:fI}) or (\ref{eq:fII}) by summing on $a$ or $\l_{n+1}$, respectively.
For short, we denote the joint probability densities (\ref{eq:fI})-(\ref{eq:fIII}) by
\beqa\label{eq:bref}
\fI(t;\l_1,\ldots,\l_n,a,n)&=&f_{\vec{\tau},A_t,N_t}(t;\l_1,\ldots,\l_n,a,n),
\nonumber\\
\fII(t;\l_1,\ldots,\l_{n+1},n)&=&f_{\vec{\tau},\tau_{N+1},N_t}(t;\l_1,\ldots,\l_{n+1},n),
\nonumber\\
\fIII(t;\l_1,\ldots,\l_n,n)&=&f_{\vec{\tau},N_t}(t;\l_1,\ldots,\l_n,n).
\eeqa
The explicit dependence in time $t$, which acts as a parameter in these densities, is enhanced by the notations above.

\subsection{Longest interval}
\label{sec:longest}

To each of these sequences corresponds a longest interval, denoted by 
\beqa\label{eq:def_lmax}
\LI(t)&=&\max(\tau_1, \tau_2, \ldots, \tau_{N}, A_t),\nonumber\\
\LII(t)&=&\max(\tau_1, \tau_2, \ldots, \tau_{N}, \tau_{N+1}),\nonumber \\
\LIII(t)&=&\max(\tau_1, \tau_2, \ldots, \tau_{N}).
\eeqa
It turns out that the ratios
\beq\label{eq:notationRV}
R^{\alpha}=\lim_{t\to\infty}\frac{\tau^{\alpha}_{\max}(t)}{t},\quad V^{\alpha}=\frac{1}{R^{\alpha}},
\qquad (\alpha= {\rm I,\, II,\, III})
\eeq
have limiting distributions, whose densities are denoted by 
$f^{\alpha}_R(r)$ and $f^{\alpha}_V(v)$.
Explicit expressions for the Laplace transforms with respect to $v$ of the latter, with $x$ conjugate to $v$, are as follows:
\beqa\label{eq:fIx}
\widehat{\fI_V}(x)=\frac{1}{1+x^\theta\e^x\int_0^x{\rm d}u\,u^{-\theta}\e^{-u}}= \frac{1}{_1 F_1(1,1-\theta,x)},
\\
\label{eq:fIIx}
\widehat{\fII_V}(x)=\e^x\widehat{\fI_V}(x),\\
\label{eq:fIIIx}
\widehat{\fIII_V}(x)=1-x^\theta\Gamma(1-\theta)\widehat{\fII_V}(x),
\eeqa
where $_1 F_1(1,1-\theta,x)$ is a confluent hypergeometric function, simply related to 
the incomplete gamma function
\beq
\Gamma(\theta,x)=\int_x^\infty{\rm d}u\,u^{\theta-1}\e^{-u},
\eeq
as follows,
\beq
_1 F_1(1,1-\theta,x)=\e^x x^\theta\left[\Gamma(1-\theta)+\theta\Gamma(-\theta,x)\right].
\eeq
The functions $\fII_V(v)$ and $\fIII_V(v)$ encountered in (\ref{eq:rosen2}) and (\ref{eq:deffVIII}), respectively, are precisely the densities of the random variables $V^{{\rm II}}$ and $V^{{\rm III}}$ defined for the second and third sequences (with $\theta=1/2$).

The expression of the Laplace transform of the density (\ref{eq:fIx}) was originally found in \cite{lamperti}, then derived by another method in \cite{gms2015}, which also addresses the same question for the two other sequences II and III.
Related studies can also be found in \cite{gms2014,ziff,gms2009} in the context of record statistics of random walks and renewal processes.

\section{The tied-down renewal process}
\label{sec:tied-down-renew}

The tied-down renewal process is defined by
the condition $\{t_{N}=t\}$, or equivalently by the condition $\{A_t=0\}$, which both express that the $N_t-$th event occurred exactly at time $t$.
This process generalizes the tied-down random walk of section \ref{sec:tiedRW}.
The stable process of order $0<\theta<1$ considered in \cite{wendel} corresponds to the tied-down renewal process considered here when the tail index of the density $\rho(\tau)$ is $0<\theta<1$. 
The results of section \ref{sec:tiedRW} on the longest interval of the Brownian bridge are recovered below by taking $\theta=1/2$.

%
\subsection{The tied-down conditional density}

The tied-down conditional density, denoted for short by
\beq\label{eq:fstar1}
f^{\star}(t;\l_1,\dots,\l_n,n)
\equiv f_{\vec{\tau},N_t|t_N}(t;\l_1,\dots,\l_n,n|\yeqt),
\eeq
is a generalization of (\ref{eq:Pstar}) (see \ref{sec:tied-down} for more details).
Its expression is
\beq\label{eq:fstar+}
f^{\star}(t;\l_1,\dots,\l_n,n)
=\frac{\rho(\l_1)\dots\rho(\l_n)\delta\left(\sum\l_i-t\right)}{\deno},
\eeq
where the denominator is obtained from the numerator by integration on the $\l_i$ and summation on $n$,
\beqa\label{eq:ftNdef+}
\deno
&=&\sum_{n\ge0}\int_{0}^{\infty}\d\l_1\dots\d\l_n\,\rho(\l_1)\dots\rho(\l_n)\delta\Big(\sum_{i=1}^n\l_i-t\Big)
\nonumber\\
&=&\sum_{n\ge0}f_{t_n}(t),
\eeqa
denoting by $f_{t_n}(t)$ the density of the sum $t_n=\tau_1+\cdots+\tau_n$, with $n$ fixed (compare to (\ref{eq:probtm}) for the discrete case).
The quantity (\ref{eq:ftNdef+}), which is the continuum counterpart of the probability $u_N$ of section \ref{sec:wendel} (see (\ref{eq:uN-renew}) and (\ref{eq:uNbis})),
is the edge value of the probability density of $t_N$ at its maximal value $t_N=t$ (see (\ref{eq:ftNdef}) and (\ref{eq:edge2})).
In Laplace space with respect to $t$, we have
\beq
\lap{t}f_{t_n}(t)=\hat\rho(s)^n,
\eeq
so
\beq\label{eq:lap-ftN}
\lap{t}\deno=\sum_{n\ge0}\hat\rho(s)^n=\frac{1}{1-\hat\rho(s)},
\eeq
which is the counterpart of (\ref{eq:uzfz}).
The right side behaves, when $s$ is small, as $s^{-\theta}/a$.
Thus, at long times, we finally obtain, using (\ref{eq:a}),
\beq\label{eq:denom}
\deno\approx 
\frac{\sin \pi\theta}{\pi}\frac{t^{\theta-1}}{\tau_0^\theta }.
\eeq
Equations (\ref{eq:fstar+})-(\ref{eq:denom}) are the cornerstones of the present study.

\subsection{Number of renewals between $0$ and $t$}
Let us consider the conditional distribution of $N_t$, the number of renewals between $0$ and $t$, for the tied-down renewal process,
\beq\label{eq:pnstar}
p^{\star}_n(t)=\prob(N_t=n|t_N=t)=
\frac{f_{t_n}(t)}{\deno},
\eeq
whose discrete counterpart is (\ref{eq:pstarm}).
We have
\beq
\langle N_t|t_N=t\rangle=\sum_{n>0}n p^{\star}_n(t)=\frac{\sum_{n>0}n f_{t_n}(t)}{\deno}.
\eeq
In Laplace space we have
\beq\label{eq:averageNt}
\lap{t}\sum_{n>0}n f_{t_n}(t)=\frac{\hat\rho(s)}{(1-\hat\rho(s))^2}\approx \frac{1}{a^2 s^{2\theta}}.
\eeq
Laplace inverting back and dividing by (\ref{eq:denom}), we obtain, at large times
\beq\label{eq:consNt}
\langle N_t|t_N=t\rangle\approx A^{\star}(\theta)\left(\frac{t}{\tau_0}\right)^\theta,\quad A^{\star}(\theta)=\frac{\Gamma(\theta)}{\Gamma(1-\theta)\Gamma(2\theta)}.
\eeq
By comparison, for the unconstrained renewal process \cite{gl2001},
\beq\label{eq:unconsNt}
\langle N_t\rangle\approx A(\theta)\left(\frac{t}{\tau_0}\right)^\theta,\quad A(\theta)=\frac{\sin\pi\theta}{\pi\theta}.
\eeq
Note that $A^{\star}(\theta)>A(\theta)$. 

Likewise
\beq\label{eq:Nt2}
\lap{t}\sum_{n\>0}n^2 f_{t_n}(t)=\frac{\hat\rho(s)(1+\hat\rho(s))}{(1-\hat\rho(s))^3}\approx \frac{2}{a^3 s^{3\theta}}.
\eeq
By inversion and division by (\ref{eq:denom}), we obtain $\langle N_t^2|t_N=t\rangle\sim t^{2\theta}$.
As for the unconstrained case \cite{gl2001}, we can set
\beq
N_t=\left(\frac{t}{\tau_0}\right)^\theta Y_t,
\eeq
where the random variable $Y_t$ has a limiting distribution when $t\to\infty$.
For instance, for $\theta=1/2$, we obtain
\beq
f_Y(y)=\frac{\pi}{2}y\,\e^{-\pi y^2/4}, \qquad y=\frac{n}{\sqrt{t/\tau_0}},
\eeq
which is the counterpart of (\ref{eq:pmN}).
More generally, we have
\beq
f_Y(y)=\frac{\pi}{\sin\pi\theta}\int\frac{{\rm d}z}{2\pi{\rm i}}\e^z\e^{-\Gamma(1-\theta)yz^\theta}.
\eeq
Setting $u=y^{1/\theta}z$ allows to relate the density of $Y$ to that of the one-sided stable law of index $\theta$.

\subsection{Marginal statistics of a single interval}
We want to determine the tied-down conditional average of one of the $\tau_i$, say $\tau_1$,
\beq
\langle\tau_1|t_N=t\rangle=
\sum_{n\ge0}\int_0^\infty\d\l_1\dots\d\l_n\,\l_1
f^{\star}(t;\l_1,\dots,\l_n,n).
\eeq
Laplace transforming the numerator of the right side yields
\beq
-\frac{\d\hat\rho(s)}{{\rm d}s}\frac{1}{1-\hat\rho(s)}\approx \frac{\theta}{s}.
\eeq
By Laplace inverting and dividing by (\ref{eq:denom}), we obtain
\beq\label{eq:single}
\langle\tau_1|t_N=t\rangle\approx B^{\star}(\theta)\tau_0^\theta t^{1-\theta},\quad B^{\star}(\theta)=\frac{\pi\theta}{\sin\pi\theta},
\eeq
which turns out to be equal to $t/\langle N_t\rangle$.
By comparison, for the unconstrained renewal process \cite{gl2001},
\beq
\langle\tau_1\rangle\approx B(\theta)\tau_0^\theta t^{1-\theta},\quad B(\theta)=\frac{\theta}{1-\theta}.
\eeq
We see that $\langle N_t|t_N=t\rangle \langle\tau_1|t_N=t\rangle$ is proportional to $t$, as expected.
Here again, the amplitude of the tied-down case $B^{\star}(\theta)$ is larger than the amplitude of the unconstrained case $B(\theta)$. 

\subsection{The longest interval}

Let $\tau^{\star}_{\max}(t)$ be the longest interval of the sequence $\tau_1,\dots,\tau_N$ with the condition that their sum $t_N=t$.
We want to compute the conditional distribution function
\beqa
F^{\star}(t;\l)
=\prob(\tau^{\star}_{\max}(t)\le\l|t_{N}=t)
\nonumber\\
\fl =\sum_{n\ge0}\int_{0}^{\l}{\rm d}\l_1\dots \int_{0}^{\l}{\rm d}\l_n f^{\star}(t;\l_1\dots,\l_n,n,\yeqt)
=\frac{F(t;\l)}{\deno},
\label{eq:defFtl}
\eeqa
where the numerator is 
\beq\label{eq:Ftljoint}
F(t;\l)=\sum_{n\ge0}\int_{0}^{\l}{\rm d}\l_1\,\rho(\l_1)\dots \int_{0}^{\l}{\rm d}\l_n\,\rho(\l_n)
\delta\Big(\sum_{i=0}^n\l_i-t\Big).
\eeq
Equation (\ref{eq:defFtl}) is the continuum counterpart of (\ref{eq:problongest}), with $F(t;\l)$ playing the role of $v_{N,k}$.
Laplace transforming (\ref{eq:Ftljoint}) with respect to time, we get
\beqa\label{eq:F}
\lap{t} F(t;\l)
=\sum_{n\ge0}\left(\int_{0}^{\l}{\rm d}\l_1\,\rho(\l_1)\e^{-s\l_1}\right)^n
=\frac{1}{1-\hat\rho(s;\l)},
\eeqa
where
\beq
\hat\rho(s;\l)=\int_{0}^{\l}{\rm d}\l_1\,\rho(\l_1)\e^{-s\l_1}.
\eeq
The expression (\ref{eq:F}) is the continuum counterpart of $\tilde v_k(z)$ given in (\ref{eq:gf}).
It holds for any distribution of intervals $\rho(\tau)$.
In the limit $\l\to\infty$ the right side is equal to $1/(1-\hat\rho(s))$, as it should (see (\ref{eq:lap-ftN})).

We now perform the asymptotic analysis of (\ref{eq:F}) along the lines of \cite{gms2015}.
An integration by parts yields
\beq
1-\hat\rho(s;\l)=p_0(\l)\e^{-s\l}+s\,\hat p_0(s;\l),
\eeq
where $\hat p_0(s;\l)=\int_{0}^{\l}\d\tau\,p_0(\tau)\e^{-s\tau}$.
Then, using the asymptotic estimates in the regime $s\to0$, $\l\to\infty$, with $s\l$ fixed \cite{gms2015},
\beq\label{eq:asym1}
\hat p_0(s;\l)\approx\tau_0^{\theta}s^{\theta-1}\int_{0}^{s\l}{\rm d}u\,u^{-\theta}\e^{-u},
\eeq
and
\beq\label{eq:asym2}
1-\hat\rho(s;\l)\approx\tau_0^\theta s^\theta\left((s\l)^{-\theta}\e^{-s\l}+\int_0^{s\l}{\rm d}u\, u^{-\theta}\e^{-u}\right),
\eeq
we obtain
\beq
\lap{t} F(t;\l)\approx
\left(\frac{\l}{\tau_0}\right)^\theta\frac{\e^{s\l}}{1+(s\l)^{\theta}\e^{s\l}\int_0^{s\l}{\rm d}u\,u^{-\theta}\e^{-u}}.
\eeq
Laplace inverting with respect to $s$ and dividing by $t^{\theta-1}\sin\pi\theta/(\pi\tau_0^\theta)$ (given in (\ref{eq:denom})), we finally obtain, using notations akin to (\ref{eq:notationRV}),
\beqa\label{eq:res1}
F^{\star}_R(r)=\overline{F}^{\star}_V(v)&=&\lim_{t\to\infty}F^{\star}(t;\l),
\nonumber\\
&=&\frac{\pi}{\sin\pi\theta} v^{1-\theta}\fII_V(v),
\eeqa
where $\fII_V(v)$ is the inverse Laplace transform of (\ref{eq:fIIx}).
Note that the microscopic scale $\tau_0$ altogether disappeared in (\ref{eq:res1}).
This means that $F^{\star}_R(r)$ is universal with respect to the choice of distribution $\rho(\tau)$, in the sense that only the tail exponent of this distribution matters.
In particular the result (\ref{eq:rosen2}), valid for the Brownian bridge, and recovered for $\theta=1/2$, is universal.
Equation (\ref{eq:res1}) can be alternatively written as
\beqa
\lap{v}v^{\theta-1}\overline{F}^{\star}_V(v)=\frac{\pi}{\sin\pi\theta}\widehat{\fII_V}(x),
\\
\lap{v}v^{\theta-1}F^{\star}_V(v)=\frac{\Gamma(\theta)}{x^\theta}\widehat{\fIII_V}(x),
\label{eq:res2}
\eeqa
using (\ref{eq:fIIIx}).
This expression, as well as its generalizations to the case of the second ($k=2$), third ($k=3$), ..., longest intervals, are given in \cite{wendel} (see $\S$ 5, theorem 4),
\beq\label{eq:orderstat}
\lap{v}v^{\theta-1}F^{(k)\star}_{V}(v)=\frac{\Gamma(\theta)}{x^\theta}\left(\widehat{\fIII_V}(x)\right)^k.
\eeq
Note that $\Gamma(\theta)/x^\theta$ is equal to the Laplace transform of $v^{\theta-1}$.
Ref \cite{wendel} also gives the generalization of the density (\ref{eq:fIx}) for the $k-$longest interval,
\beq\label{eq:orderstat-freeI}
\widehat{f^{(k){\rm I}}_{V}}(x)=\widehat{\fI_V}(x)\left(\widehat{\fIII_V}(x)\right)^{k-1}.
\eeq
Equations (\ref{eq:orderstat}) and (\ref{eq:orderstat-freeI}) are derived in \ref{app:orderstat} by elementary methods of order statistics theory.
Likewise, one could show that
\beq\label{eq:orderstat-freeII}
\widehat{f^{(k){\rm II}}_{V}}(x)=\widehat{\fII_V}(x)\left(\widehat{\fIII_V}(x)\right)^{k-1},
\eeq
and
\beq\label{eq:orderstat-freeIII}
\widehat{f^{(k){\rm III}}_{V}}(x)=\left(\widehat{\fIII_V}(x)\right)^{k},
\eeq
which can be summarized as
\beq\label{eq:orderstat-freex}
\widehat{f^{(k)\alpha}_{V}}(x)=\widehat{f^\alpha_V}(x)\left(\widehat{\fIII_V}(x)\right)^{k-1},\quad (\alpha={\rm I, II, III}),
\eeq
with the definitions (\ref{eq:fIx})-(\ref{eq:fIIIx}).

\subsection{Average longest interval}

The method follows that of section (\ref{sec:averageRW}).
The average longest interval is computed as
\beq\label{eq:taumax}
\left\langle\tau^{\star}_{\max}(t)\right\rangle=\int_0^\infty{\rm d}\l\, \overline{F}^{\star}(t;\l).
\eeq
We have (see (\ref{eq:defFtl}))
\beq
F(t;\l)+\overline{F}(t;\l)=\deno.
\eeq
Laplace transforming this equation with respect to time, we get
\beqa\label{eq:lapFbar}
\lap{t}\overline{F}(t;\l)
&=&\frac{1}{1-\hat\rho(s)}-\frac{1}{1-\hat\rho(s;\l)}
\\
&\approx&\frac{s^{-\theta}}{a}\left(1-(s\l)^\theta\Gamma(1-\theta)\widehat{\fII_V}(s\l)\right)
=\frac{s^{-\theta}}{a}\widehat{\fIII_V}(s\l).
\eeqa
After integration upon $\l$, inverse Laplace transform with respect to $s$, and division by $\deno$ (given by (\ref{eq:denom})), we obtain
\beqa\label{eq:Rmoyen}
\langle R^{\star}\rangle&=&\lim_{t\to\infty}\left\langle\frac{\tau^{\star}_{\max}(t)}{t}\right\rangle
=\frac{1}{\theta}\int_0^\infty{\rm d}x\,\widehat{\fIII_V}(x)
\\
&=&\frac{1}{\theta}\lim_{t\to\infty}\left\langle\frac{\LIII(t)}{t}\right\rangle
=\frac{1}{\theta}\langle R^{\rm III}\rangle.
\eeqa
For $\theta=1/2$ we recover (\ref{eq:averageRW}).

Using the same method, we find, for the $k-$th longest interval,
\beqa\label{eq:Rmoyenk}
\langle R^{(k)\star}\rangle=\lim_{t\to\infty}\left\langle\frac{\tau^{(k)\star}_{\max}(t)}{t}\right\rangle
&=&\frac{1}{\theta}\int_0^\infty{\rm d}x\,\left(\widehat{\fIII_V}(x)\right)^k
\nonumber
\\
&=&\frac{1}{\theta}\langle R^{(k){\rm III}}\rangle.
\eeqa
Since the sum of the intervals $\tau^{(k)\star}_{\max}(t)$ is, by definition of the process, equal to $t$, one should 
have
\beq
\sum_{k\ge1}\langle R^{(k)\star}\rangle=1.
\eeq
This result can indeed be proved by direct computation of the integral of the geometrical series in $\widehat{\fIII_V}(x)$ in the right side, which gives the value $\theta$.
One can also note that the averages $\langle R^{(k){\rm III}}\rangle$ sum up to $\langle t_N/t\rangle$ and that 
$\langle t_N/t\rangle\to\theta$ \cite{gl2001}.
(See also \cite{veto,bar}.)

\subsubsection*{Remark} The last equality in (\ref{eq:Rmoyenk}) can be obtained by the methods of \ref{app:orderstat} (see \cite{veto}).
A list of values of $\langle R^{(k){\rm III}}\rangle$ for $\theta=1/2$ can be found in \cite{veto}.
One could generalize the results of \cite{veto} to find
\beq
\langle R^{(k){\rm I}}\rangle
=\int_0^\infty{\rm d}x\,\widehat{f^{(k){\rm I}}_{V}}(x),\quad \langle R^{(k){\rm II}}\rangle
=\int_0^\infty{\rm d}x\,\widehat{f^{(k){\rm II}}_{V}}(x).
\eeq

\subsection{Characterization of the densities $f^{\star}_V$ and $f^{\star}_R$ }
\label{sec:charact}

The denominator of $\widehat{\fII_V}(x)$ in (\ref{eq:fIx}), $D(x)
=1+x^\theta\e^x\int_0^x{\rm d}u\,u^{-\theta}\e^{-u}$, satisfies the following differential equation 
\beq\label{eq:eqdiff}
xD'(x)=D(x)(x+\theta)-\theta.
\eeq
The residues at the poles $x_k$ of $\widehat{\fII_V}(x)$ are therefore equal to $-x_k\e^{x_k}/\theta$, so
\beq
F_R^{\star}(r)=
\overline{F}_V^{\star}(v)
=\frac{\pi}{\theta\sin\pi\theta}v^{1-\theta}\sum_{k=-\infty}^{\infty}(-x_k)\e^{x_k(1+v)},
\eeq
which generalizes (\ref{eq:rosen1}) \cite{wendel}.
Hence the asymptotic behaviour at large $v$ (small $r$) of the corresponding densities is obtained as in section \ref{sec:wendelfstar} (see (\ref{eq:fVas}), (\ref{eq:fRas})), yielding
\beqa
f^{\star}_V(v)\approx \frac{\pi x_0^2}{\theta\sin\pi\theta} v^{1-\theta}\,\e^{-|x_0|(1+v)}\left(1-\frac{1-\theta}{|x_0|v}\right),\quad 
\nonumber\\
f^{\star}_R(r)\approx \frac{\pi\, x_0^2}{\theta\sin\pi\theta}\frac{\e^{-|x_0|(1
+\frac{1}{r})}}{r^{3-\theta}}\left(1-\frac{(1-\theta)r}{|x_0|}\right).
\eeqa
These expressions only depend on the tail index $\theta$ and therefore are universal.
The density $f_R^{\star}(r)$ for $1/2<r<1$ has a simple expression,
\beq\label{eq:fr12}
f_R^{\star}(r)=\frac{\theta\,\Gamma(\theta)^2\sin \pi\theta }{\pi\Gamma(2\theta)}\frac{(1-r)^{2\theta-1}}{r^{1+\theta}},
\eeq
which, for $\theta=1/2$, yields back $f_R^{\star}(r)=r^{-3/2}/2$.
This expression is obtained by the following reasoning, adapted from that used in \cite{wendel} for the tied-down random walk, and reproduced in section \ref{sec:wendelfstar} (see (\ref{eq:rosenbis})).
Let $w(t;\l)$ be the density 
\beq
w(t;\l)=\frac{{\rm d}F(t;\l)}{{\rm d}\l},
\eeq
where $F(t;\l)$ is defined in (\ref{eq:Ftljoint}).
If $\l>t/2$, 
then on can 
decompose an history into three contributions as in (\ref{eq:rosenbis}), yielding
\beq
w(t;\l)=\int_{0}^{t-\l}{\rm d}\tau\, U(\tau)\rho(\l)U(t-\l-\tau).
\eeq
Noting that
\beq
\lap{T}\int_{0}^{T}{\rm d}\tau\, U(\tau)U(T-\tau)=\frac{1}{(1-\hat\rho(s))^2}\approx (as)^{-2\theta},
\eeq
we have, for $\l>t/2$,
\beq
w(t;\l)\approx \frac{1}{a^{2\theta}\Gamma(2\theta)}\rho(\l)(t-\l)^{2\theta-1}.
\eeq
It follows that
\beq\label{eq:FRstarr}
F_R^{\star}(r)=1-\frac{C(\theta)}{t^{\theta-1}}\int_{rt}^{t}\d\l\,\l^{-1-\theta}(t-\l)^{2\theta-1},\quad (1/2<r<1),
\eeq
where
\beq
C(\theta)=\frac{\theta\,\Gamma(\theta)^2\sin \pi\theta }{\pi\Gamma(2\theta)},
\eeq
yielding (\ref{eq:fr12}), by derivation of (\ref{eq:FRstarr}) with respect to $r$.
As a consequence
\beq
f_{V}^{\star}(v)=\frac{\theta\,\Gamma(\theta)^2\sin \pi\theta }{\pi\Gamma(2\theta)}
v^{-\theta}(v-1)^{2\theta-1},\quad (1<v<2).
\eeq
It is also possible to derive (\ref{eq:fr12}) from equation (6.1) of \cite{wendel}.

\subsection{Probability of record breaking $\Q(t)$}
\label{sec:Qt}

The probability that the last interval is the longest one is defined as in section \ref{sec:breaking},
\beq
\fl\Q(t)
=\prob(\tau^{\star}_{\max}(t)=\tau_{N}|t_N=t)
=\prob(\tau_{N}>\max(\tau_1,\ldots,\tau_{N-1})|t_N=t).
\eeq
This probability is given by the sum (see \cite{gms2015} for similar reasonings)
\beq
\fl\Q(t)=\sum_{n\ge1}Q^{\star}_n(t)=\sum_{n\ge1}\prob\Big(\tau_{N}>\max(\tau_1,\ldots,\tau_{N-1}),N_t=n|t_N=t\Big).
\eeq
Explicitly,
\beqa
\fl \Q_n(t)=\int_0^\infty{\rm d}\l_{n}\,\int_0^{\l_n}{\rm d}\l_1\,\ldots\int_0^{\l_n}{\rm d}\l_{n-1}\,
f^{\star}(t;\l_1,\dots,\l_{n-1},\l_{n},n)
\\
=\frac{Q_n(t)}{\deno},
\eeqa
where
\beq
\fl Q_n(t)=\int_0^\infty{\rm d}\l_{n}\,\int_0^{\l_n}{\rm d}\l_1\,\ldots\int_0^{\l_n}{\rm d}\l_{n-1}\,
\rho(\l_1)\dots\rho(\l_{n})\,\delta\Big(\sum_{i=0}^n\l_i-t\Big).
\eeq
In Laplace space, after summing on $n$, we have
\beq
\hat Q(s)
=\int_0^\infty{\rm d}\l\,\frac{\rho(\l)\e^{-s\l}}{1-\int_0^\l{\rm d}\tau\,\rho(\tau)\e^{-s\tau}}
=\int_0^{\hat\rho(s)}\frac{{\rm d}\hat\rho(s;\l)}{1-\hat\rho(s;\l)},
\eeq
where $\hat\rho(s;\l)=\int_0^\l{\rm d}\tau\,\rho(\tau)\e^{-s\tau}$.
Finally,
\beq\label{eq:Q*}
\hat Q(s)
=-\ln(1-\hat\rho(s)).
\eeq

The same result can be recovered by assuming that the $N_t$ intervals $\tau_1,\ldots,\tau_{N}$ should all play the same role, hence that the probability of record breaking is equal to the inverse number of these random variables, as for iid random variables.
So, let us assume that
\beq
Q^{\star}_n(t)=\frac{p^{\star}_n(t)}{n},\quad (n>0),
\eeq
where $p^{\star}_n(t)=\prob(N_t=n|t_N=t)$ (see (\ref{eq:pnstar})).
Thus
\beq\label{eq:Qstar}
Q^{\star}(t)=
\sum_{n\ge1}Q^{\star}_n(t)=\left\langle N_t^{-1}|t_N=t\right\rangle.
\eeq
In Laplace space, the numerator of this expression is
\beqa\label{eq:hatQs}
\hat Q(s)&=&\sum_{n\ge1}\frac{\hat f_{t_n}(s)}{n}
=\sum_{n\ge1}\frac{\hat \rho(s)^n}{n},
\nonumber\\ 
&=&-\ln(1-\hat\rho(s)),
\eeqa
which is (\ref{eq:Q*}) above.
The last step consists in Laplace inverting this expression with respect to $s$, then dividing by $\deno$.
We thus find, at large times,
\beq\label{eq:Qtrenew}
\Q(t)\approx \frac{\pi\theta}{\sin\pi\theta}\left(\frac{\tau_0}{t}\right)^\theta\approx\frac{1}{\langle N_t\rangle},
\eeq
where the right side pertains to the unconstrained case (see (\ref{eq:unconsNt})).
There is no universality of the result with respect to the choice of distribution $\rho(\tau)$ since the microscopic scale $\tau_0$ is still present (compare to (\ref{eq:QN+})).
We also recall, for comparison, that $\QIII(t)\sim \ln t/t^\theta$ \cite{gms2015}.

\subsection{Narrow distribution of intervals}

The aim of this subsection is to determine the distribution of $\tau^{\star}_{\max}(t)$ and the probability of record breaking $\Q(t)$ for a narrow distribution of intervals, taking
the exponential distribution of intervals, $\rho(\tau)=\e^{-\tau}$, as an example.
We first note that, by inversion of (\ref{eq:lap-ftN}), we have $\deno=1$ for $t>0$.

The computation of $\left\langle\tau^{\star}_{\max}(t)\right\rangle$ relies on (\ref{eq:taumax}) and (\ref{eq:lapFbar}).
We find
\beq
\int_0^{\infty}\d\l\,\lap{t}\overline{F}(t;\l)=\frac{1}{s}\ln\left(1+\frac{1}{s}\right),
\eeq
whose inverse is
\beq
\int_0^{\infty}\d\l\,\overline{F}(t;\l)={\rm E}(t)\equiv\int_0^t{\rm d}u\,\frac{1-\e^{-u}}{u}.
\eeq
At large times, ${\rm E}(t)\approx\ln t+\gamma$, where $\gamma$ is the Euler constant.
We finally obtain
\beq
\left\langle\tau^{\star}_{\max}(t)\right\rangle={\rm E}(t)\approx\ln t+\gamma.
\eeq
We also have, for $\e^{-\l}\sim s\ll1$,
\beq
\lap{t}F(t;\l)\approx\frac{1}{s+\e^{-\l}},
\eeq
which by inversion yields
\beq
F(t;\l)=F^{\star}(t;\l)\approx \e^{-\e^{-(\l-\ln t)}}.
\eeq
So
\beq
\tau^{\star}_{\max}(t)\approx \ln t+Z^{\rm G},
\eeq
where $Z^{\rm G}$ follows the standard Gumbel distribution, with $\langle Z^{\rm G}\rangle=\gamma$.
This behaviour coincides with that found for the three sequences (\ref{eq:sequences}) in \cite{gms2015}.
We also find, from (\ref{eq:hatQs}) that $Q^{\star}(t)\approx 1/t$.
Since $\langle N_t|t_N=t\rangle\approx t$, as can be inferred from (\ref{eq:averageNt}),
$Q^{\star}(t)$ behaves qualitatively as if the $N_t$ intervals were iid random variables.
This is akin to what was found for the cases of $\QI(t)$ and $\QIII(t)$, the probabilities of record breaking for the sequences I and III \cite{gms2015}.

\subsection{Broad distribution of intervals with $\theta>1$ }

We first find, by inversion of (\ref{eq:lap-ftN}), that 
\beq\label{eq:ftN2}
\deno\approx \frac{1}{\langle\tau\rangle}+\frac{\tau_0^\theta}{(\theta-1)\langle\tau\rangle^2}t^{1-\theta}.
\eeq
We then compute the average number of renewals, using the first equality in (\ref{eq:averageNt}).
We obtain, after division by $1/\langle\tau\rangle$,
\beq
\langle N_t|t_N=t\rangle\approx \frac{t}{\langle\tau\rangle}+
\frac{2\tau_0^\theta}{(\theta-1)(2-\theta)\langle\tau\rangle^2}t^{2-\theta}.
\eeq

We restart from (\ref{eq:F}) in order to compute the distribution of $\tau^{\star}_{\max}(t)$.
Following the asymptotic analysis made in \cite{gms2015}, we find
\beq
\lap{t}F(t;\l)\approx\frac{1}{\langle\tau\rangle}\frac{1}{s+(\l/\tau_0)^{-\theta}/\langle\tau\rangle},
\eeq
hence
\beq
F(t;\l)\approx \frac{1}{\langle\tau\rangle}\e^{-t/\langle\tau\rangle(\l/\tau_0)^{-\theta}}.
\eeq
Dividing this expression by the leading order $1/\langle\tau\rangle$ in (\ref{eq:ftN2}), we have
\beq
F^{\star}(t;\l)\approx \e^{-t/\langle\tau\rangle(\l/\tau_0)^{-\theta}}.
\eeq
Setting
\beq
\tau^{\star}_{\max}(t)=\tau_0\left(\frac{t}{\langle\tau\rangle}\right)^{1/\theta}Z_t,
\eeq
we have, as $t\to\infty$, $Z_t\to Z^{F}$, with limiting distribution 
\beq
\prob(Z^{F}<x)=\e^{-1/x^\theta} \label{eq:frechet}
\eeq
which is the Fr\'echet law.
Therefore
\beq\label{LI>1}
\langle\tau^{\star}_{\max}(t)\rangle\approx\tau_0\left(\frac{t}{\langle\tau\rangle}\right)^{1/\theta}\underbrace{\langle Z^{F}\rangle}_{\Gamma(1-1/\theta)}.
\eeq
This is exactly the result found for the three sequences (\ref{eq:def_lmax}) in \cite{gms2015}.
The tied-down condition does not change the asymptotic distribution of the longest interval if $\theta>1$.
Finally, from (\ref{eq:hatQs}) we find
\beq
Q^{\star}(t)\approx \frac{\langle\tau\rangle}{t},
\eeq
which has therefore the same time dependence as $\QIII(t)$ \cite{gms2015}.

\section{Summary and discussion}

The tied-down renewal process studied in the present paper is equivalent to the stable process considered in \cite{wendel} if the tail exponent of the distribution of intervals is comprised between 0 and 1.
The results of \cite{wendel} concerning the statistics of the longest interval are thus recovered in a simple manner.
The method consists in considering the joint probability (\ref{eq:Pstar}) for the case of the random walk or the joint probability density (\ref{eq:fstar+}) for the case of a continuous renewal process.

\begin{table}[h]
\caption{Some important results for the tied-down random walk of section \ref{sec:tiedRW} (starting and ending at the origin) and the continuum renewal process of section \ref{sec:tied-down-renew} (with $0<\theta<1$).
The results in the left column correspond, respectively, to the mean number of intervals
(\ref{eq:meanMN}), the asymptotic mean ratio of the longest ranked interval to the total length of time (generalizing (\ref{eq:averageRW})), and the probability of record breaking (\ref{eq:QN+}).
Those in the right column correspond to the equivalent quantities
(\ref{eq:consNt}), (\ref{eq:Rmoyenk}), and (\ref{eq:Qtrenew}), in the continuum formalism.
}
\label{tab:table2}
\begin{center}
\begin{tabular}{|c|c|}
\hline
 tied-down random walk & tied-down renewal proc. \\
\hline
$\langle M_N|S_{2N}=0\rangle\approx\sqrt{\pi N}$&$\langle{N_t|t_N=t}\rangle\approx\frac{\Gamma(\theta)}{\Gamma(1-\theta)\Gamma(2\theta)}
\left(\frac{t}{\tau_0}\right)^\theta$\\
$\langle{R^{(k)\star}}\rangle=2\langle{R^{(k){\rm III}}}\rangle$&$\langle{R^{(k)\star}}\rangle=\frac{1}{\theta}\langle{R^{(k){\rm III}}}\rangle$\\
$Q_N^{\star}\approx\frac{\sqrt{\pi}}{2\sqrt{N}}$&$Q^{\star}(t)\approx\frac{\pi\theta}{\sin\pi\theta}\left(\frac{\tau_0}{t}\right)^\theta$\\
\hline
\end{tabular}
\end{center}
\end{table}

This study is extended in several directions such as the statistics of the number of intervals or the probability of record breaking, both for the tied-down random walk and the tied-down renewal process.
We also discuss the cases of a narrow distribution of intervals or of a distribution with a tail exponent $\theta>1$.
A summary of some important results for the tied-down random walk and the tied-down renewal process (for $0<\theta<1$) is given in table \ref{tab:table2}.
The results obtained in section \ref{sec:tiedRW} for the Brownian bridge
are recovered in the formalism of the tied-down renewal process by taking $\theta=1/2$.

Some related works, that we now review, can be put in perspective with the present study.
First, as emphasized in the course of this study, there are close connections between the tied-down renewal process, including the Brownian bridge, and the cases (I, II and III) considered in \cite{gms2015,gms2014,veto}.
In particular, several connections with case III were emphasized in the previous sections (see also table \ref{tab:table2}).

Then, in the recent past, the distribution of the longest interval for the tied-down random walk and the Brownian bridge of section \ref{sec:tiedRW} was investigated in \cite{fik}.
The results of the analysis performed in \cite{fik} can be usefully completed by the studies made in \cite{wendel,pitman,lindell} and in the present work.

Finally, a recent study of the largest domain in a specific one-dimensional system of Ising spins has been given in \cite{bar}.
In this model, introduced in \cite{bar1,bar2}, the Boltzmann weight of a spin configuration can be expressed in terms of the lengths of domains.
The parameters defining the model are the temperature and the exponent characterizing the power-law decay of the distribution of the lengths of the domains (denoted by $\theta$ in the present work).
When the spin system is at criticality, this Boltzmann weight can be seen as the discrete version (\ref{eq:Pstar}) of the tied-down conditional density (\ref{eq:fstar+}), for a particular choice of distribution of intervals $f_{\l}$.
In other words, the general probabilistic framework presented here for the tied-down renewal process encompasses the analysis made in \cite{bar} for the spin system at criticality.
In particular, the results given in \cite{bar} for the statistics of the largest spin domain at criticality (with $0<\theta<1$) coincide asymptotically with their counterparts for the longest interval of the tied-down renewal process, 
as a consequence of the universality of these results with respect to the choice of distribution of intervals $\rho(\tau)$, when $0<\theta<1$, as demonstrated in the present work. 
Note that the analysis performed here for the number of intervals $M_N$ (in the discrete case), or $N_t$ (in the continuous case), provides an answer to the question of the statistics of the fluctuating number of domains in the spin model considered in \cite{bar1,bar2,bar}.

In a companion paper \cite{prepa} we will complete the study done here by addressing the statistics of other quantities, such as the occupation time or the two-time correlation function of the process.

To close, let us mention an interesting connection between the problem discussed in the present work and the statistics of records for general random walks with symmetric distributions of steps (for a review, see \cite{review}).
The analysis performed here applies to the case where the last record of the random walk is conditioned to occur at the last step $N$, or,
said differently, when the maximum of the random walk occurs exactly at the last step $N$.

\ack I thank K. Xu for pointing out a typo 
in the preliminary version of this work.
\newpage

\appendix


\section{Notations}
\label{sec:notations}

The distribution function of the random variable $X$ is denoted by
\beq
F_X(x)=\prob(X\le x).
\eeq
If $X$ is a continuous random variable, it has a density
\beq
f_X(x)=\frac{{\rm d}F_X(x)}{{\rm d}x}.
\eeq
For several random variables we have
\beq
F_{X_1,X_2,\ldots}(x_1,x_2,\ldots)=\prob(X_1\le x_1,X_2\le x_2,\ldots),
\eeq
with associated density $f_{X_1,X_2,\ldots}(x_1,x_2,\ldots)$.
When permitted by the context, we will omit the variables in subscript.

Let $X$ and $Y$ two random variables with joint density $f_{X,Y}(x,y)$ and marginal densities $f_X(x)$ and $f_Y(y)$.
For discrete random variables the conditional distribution function of $X$ given $Y=y$ is simply
\beq
\prob(X\le x|Y=y)=F_{X|Y}(x|y)=\frac{\prob(X\le x,Y=y)}{\prob(Y=y)}.
\eeq
For continuous random variables, the conditional distribution function of $X$ given $Y=y$ is defined as follows \cite{grimmett},
\beq
\prob(X\le x|Y=y)=F_{X|Y}(x|y)=\int_0^x {\rm d}u\,\frac{f_{X,Y}(u,y)}{f_Y(y)}.
\eeq
Therefore the conditional density reads
\beq\label{app:conditional}
f_{X|Y}(x|y)=\frac{f_{X,Y}(x,y)}{f_Y(y)}=\frac{f_{X,Y}(x,y)}{\int{\rm d}x\,f_{X,Y}(x,y)}.
\eeq

\section{First return probability $f_n$ for the simple random walk}
\label{sec:return}

Let 
\beq\label{eq:return}
u_n=\prob(S_{2n}=0)=(-1)^n\stirl{-\frac{1}{2}}{ n}=\frac{1}{2^{2n}}\stirl{2n}{n},
\eeq
and
\beq\label{eq:first}
\fl f_n=\prob(\textnormal{first return to zero occurs at time } 2n)=(-1)^{n-1}\stirl{\frac{1}{2}}{n}.
\eeq
Thus,
$
 u_0=1,u_1=\frac{1}{2},u_2=\frac{3}{8}\ldots;
f_1=\frac{1}{2},f_2=\frac{1}{8},f_3=\frac{1}{16}\ldots.
$
These probabilities obey $f_n=u_{n-1}-u_n$, and their generating functions are
\beqa\label{app:utilde}
\tilde u(z)=\sum_{n\ge0}u_{n}z^n=\frac{1}{\sqrt{1-z}},
\\
\label{app:ftilde}
\tilde f(z)=\sum_{n\ge0}f_{n}z^n=1-\sqrt{1-z}.
\eeqa
At large $n$,
\beq\label{eq:asymu}
u_n\approx\frac{1}{\sqrt{\pi n}},\qquad f_n\approx\frac{1}{2\sqrt{\pi}n^{3/2}}.
\eeq

\section{The tied-down conditional density}
\label{sec:tied-down}
Let us add some details on the definition of the tied-down conditional density (\ref{eq:fstar+}) given in the bulk of the paper.
Consider the conditional probability
\beq
\prob(\vec{\tau}\le \vec{\l},N_t=n|t_N=y),
\eeq
where $\vec{\l}=\{\l_1,\dots,\l_n\}$ is a realization of the sequence of intervals
\beq
\vec{\tau}=\{\tau_1,\dots,\tau_N\}.
\eeq
The associated conditional density is a generalization of (\ref{app:conditional}), with $X=\{\vec{\tau},N_t\}$ and $Y=t_N$,
\beq\label{eq:conditional1}
f_{\vec{\tau},N_t|t_N}(\l_1,\dots,\l_n,n|y)=\frac{f_{\vec{\tau},N_t,t_N}(t;\l_1,\dots,\l_n,n,y)}{f_{t_N}(y)}.
\eeq
The numerator is explicitly obtained by multiplying $\fIII(\cdot)$ (given by (\ref{eq:fIII})) by $\delta(\sum\l_i-y)$, i.e., 
\beqa\label{eq:jointe}
f_{\vec{\tau},N_t,t_N}(t;\l_1,\dots,\l_n,n,y)
\nonumber \\
=\r(\l_1)\ldots\r(\l_n)\,\int_0^\infty{\rm d}a\,p_0(a)\,\delta\Big(\sum_{i=1}^n\l_i+a-t\Big)\delta\Big(\sum_{i=1}^n\l_i-y\Big).
\eeqa
The denominator, $f_{t_N}(t;y)$, obtained from the numerator by integration on $\l_1,\dots,\l_n$ and summation on $n$, is the probability density of the random variable $t_{N}$.
The double Laplace transforms with respect to $t$ and $y$ of the numerator and of the denominator are respectively given in (\ref{eq:lapNumerator}) 
and (\ref{eq:laptN}).

The tied-down conditional density (\ref{eq:fstar}) is defined as (\ref{eq:conditional1}), however with the condition that $t_N=t$.
Setting $y=t$ in (\ref{eq:jointe}) amounts to suppressing the first delta function in the right side of the equation.
The remaining integral upon $a$ is equal to 1, so
\beq\label{eq:J-density}
f_{\vec{\tau},N_t,t_N}(t;\l_1,\dots,\l_n,n,\yeqt)=
\rho(\l_1)\dots\rho(\l_n)\delta\Big(\sum_{i=1}^n\l_i-t\Big).
\eeq
The same result can also be obtained by setting $a=0$ in (\ref{eq:fIjoint}).
Thus, using a shorter notation for the tied-down conditional density, we have
\beqa\label{eq:fstar}
f^{\star}(t;\l_1,\dots,\l_n,n)
&=&f_{\vec{\tau},N_t|t_N}(t;\l_1,\dots,\l_n,n|\yeqt)
\nonumber\\
&=&\frac{\rho(\l_1)\dots\rho(\l_n)\delta\left(\sum\l_i-t\right)}{f_{t_N}(t;\yeqt)},
\eeqa
where the denominator reads\footnote{$n=0$ corresponds to $\delta(t)$ in (\ref{eq:J-density}), and therefore to 1 in Laplace space.}
\beqa\label{eq:ftNdef}
f_{t_N}(t;\yeqt)
&=&\sum_{n\ge0}\int_{0}^{\infty}\d\l_1\dots\d\l_n\,\rho(\l_1)\dots\rho(\l_n)\delta\Big(\sum_{i=1}^n\l_i-t\Big)
\nonumber\\
&=&\sum_{n\ge0}f_{t_n}(t),
\eeqa
denoting by $f_{t_n}(t)$ the density of the sum $t_n=\tau_1+\cdots+\tau_n$, with $n$ fixed.
In the bulk of the paper we use the shorter notation
\beq\label{eq:notationftN}
\deno\equiv f_{t_N}(t;\yeqt)
\eeq
for the edge value of the probability density of $t_N$ at its maximal value $y=t$.

\subsubsection*{Remark}

Laplace transforming (\ref{eq:jointe}) with respect to $t$ and $y$ (with $s$ conjugate to $t$ and $u$ conjugate to $y$), yields
\beqa\label{eq:lapNumerator}
\lap{t,y}f_{\vec{\tau},N_t,t_N}(t;\l_1,\dots,\l_n,n,y)
\nonumber \\
=\r(\l_1)\e^{-(s+u)\l_1}\ldots\r(\l_n)\e^{-(s+u)\l_1}\,\frac{1-\hat\rho(s)}{s}.
\eeqa
Then summing upon the $\l_i$ and $n$ yields the double Laplace transform of $f_{t_N}(t;y)$ \cite{gl2001}
\beq\label{eq:laptN}
\lap{t,y}f_{t_N}(t;y)
=\lap{t}\langle\e^{-u t_N}\rangle=\frac{1}{1-\hat\rho(s+u)}\frac{1-\hat\rho(s)}{s}.
\eeq
In order to get the edge value of this density at $y=t$, we invert (\ref{eq:laptN}),
\beq
f_{t_N}(t;y)=\int\frac{{\rm d}u}{2i\pi}\e^{uy}\int\frac{{\rm d}s}{2i\pi}\e^{st}\frac{1}{1-\hat\rho(s+u)}\frac{1-\hat\rho(s)}{s},
\eeq
we then set $\yeqt$ and $w=s+u$, yielding, with the shorter notation (\ref{eq:notationftN}),
\beq\label{eq:edge2}
\deno=\int\frac{{\rm d}w}{2i\pi}\e^{wt}\frac{1}{1-\hat\rho(w)}\int\frac{{\rm d}s}{2i\pi}\frac{1-\hat\rho(s)}{s}.
\eeq
The second integral is equal to 1, since it represents $p_0(t)$ for $t=0$.
We thus recover (\ref{eq:lap-ftN}).

\section{Second, third, \dots, longest intervals}
\label{app:orderstat}
For independent, identically distributed random variables $X_1,\dots,X_n$, the distribution function of the $k-$th largest random variable $X^{(k)}$ can be obtained by noting that the event $\{X^{(k)}\le\l\}$ means that at most $k-1$ variables $X_i$ are larger than $\l$, so
\beqa\label{app:Fkl}
F^{(k)}(\l)=\prob(X^{(k)}\le\l)
&=&\sum_{j=0}^{k-1}\prob(j \ {\rm r.\,v.}\ X_i>\l)
\\
&=&\sum_{j=0}^{k-1}\binom{n}{j}\overline{F}(\l)^{j} F(\l)^{n-j},
\eeqa
where $F(\l)=\prob(X\le\l)$, $\overline{F}(\l)=\prob(X>\l)$.

For the intervals $\tau_1,\dots,\tau_N$, the conditional distribution function
\beq
F^{(k)\star}(t;\l)=\prob(\tau_{\max}^{(k)\star}\le\l|t_N=t)
\eeq
still obeys (\ref{app:Fkl}).
We have likewise, using (\ref{eq:fstar}), 
\beqa
\fl\prob(j \ {\rm r.\,v.}\ \tau_i>\l)=
\frac{1}{\deno}
\\
\sum_{n\ge0}\binom{n}{j}\underbrace{\int_\l^\infty {\rm d}\l_1\,\rho(\l_1)\cdots}\,
\underbrace{\int_0^\l {\rm d}\l_1\,\rho(\l_1)\cdots}
\delta\Big(\sum_{i=1}^n\l_i-t\Big),
\eeqa
where the first group of integrals is done $j$ times, and the second group $n-j$ times.
Summing on $j$ and Laplace transforming with respect to time, we obtain for the numerator of $F^{(k)\star}(t;\l)$,
denoted by
\beq
F^{(k)}(t;\l)=\prob(\tau_{\max}^{(k)\star}\le\l,t_N=t),
\eeq
the expression
\beqa
\lap{t}F^{(k)}(t;\l)
=\sum_{j=0}^{k-1}\,\sum_{n\ge0}\binom{n}{j}\left[\hat\rho(s)-\hat\rho(s;\l)\right]^{j}
\left[\hat\rho(s;\l)\right]^{n-j}
\\
=\sum_{j=0}^{k-1}\,
\frac{\left[\hat\rho(s)-\hat\rho(s;\l)\right]^{j}}{\left[1-\hat\rho(s;\l)\right]^{j+1}}
=\frac{1	}{1-\hat\rho(s)}\left(1-\left[\frac{\hat\rho(s)-\hat\rho(s;\l)}{1-\hat\rho(s;\l)}\right]\right)^k.
\eeqa
In the scaling limit of large times, i.e., $s\to0$, using the asymptotic estimate (\ref{eq:asym2}), this expression becomes
\beq
\lap{t}F^{(k)}(t;\l)=
\frac{1-\left(\widehat{\fIII_V}(s\l)\right)^k}{as^\theta}.
\eeq
By Laplace inversion with respect to $s$, and division by (\ref{eq:denom}), we obtain Wendel's result (\ref{eq:orderstat}).
A similar computation is done in \cite{bar} for the case of a specific choice of discrete distribution of intervals.

We can finally derive (\ref{eq:orderstat-freeI}) by the same methods.
We start from the distribution $\fI(\cdot)$, given by (\ref{eq:fIjoint}),
for the $N_t+1$ intervals $\tau_1,\dots,\tau_N,A_t$.
In order to evaluate the probability of having $j$ of these random variables larger than $\l$, we have to separate the cases where $A_t$ belongs to the group of random variables smaller than $\l$ or to the group of random variables larger than $\l$.
Hence
\beqa
\fl\prob(j \ {\rm r.\,v.}\ (\tau_i \ {\rm and}\ A_t) >\l)=
\nonumber\\
\fl\sum_{n\ge0}\binom{n}{j-1}\underbrace{\int_\l^\infty {\rm d}\l_1\,\rho(\l_1)\cdots}\,
\underbrace{\int_0^\l {\rm d}\l_1\,\rho(\l_1)\cdots}
\int_{\l}^{\infty}{\rm d}{a}\,p_0(a)\,
\delta\Big(\sum_{i=1}^n\l_i+a-t\Big)+
\nonumber\\
\fl \sum_{n\ge0}\binom{n}{j}\underbrace{\int_\l^\infty {\rm d}\l_1\,\rho(\l_1)\cdots}\,
\underbrace{\int_0^\l {\rm d}\l_1\,\rho(\l_1)\cdots}
\int_{0}^{\l}{\rm d}{a}\,p_0(a)\,
\delta\Big(\sum_{i=1}^n\l_i+a-t\Big).
\eeqa
In the first line of the right side, the first group of integrals is done $j-1$ times, and the second group $n-j+1$ times, with $j\ge1$, while in the second line, the first group of integrals is done $j$ times, and the second group $n-j$ times, with $j\ge0$.
The rest of the computation follows as above, using the asymptotic estimates (\ref{eq:asym1}) and (\ref{eq:asym2}).

\newpage
\end{document}